\documentclass[smallextended]{svjour3}       
\smartqed  
\usepackage{graphicx}
\usepackage{mathptmx}      
\usepackage{epstopdf, epsfig}
\usepackage{amsmath,amssymb,amsfonts}
\usepackage{bm}

\newcommand{\pd}[1]{\partial_{#1}}
\newcommand{\dd}[3]{\frac{\text{d}^{#3}#1}{\text{d}#2^{#3}}}
\newcommand{\bu}{\bm{u}}
\newcommand{\eps}{\epsilon}
\newcommand{\bcdot}{\mathbf{\cdot}}
\newcommand{\bnabla}{\mathbf{\nabla}}
\newcommand{\mat}[1]{\text{\textbf{\textsf{#1}}}}
\newcommand{\imag}{\text{i}}

\journalname{Theoretical and Computational Fluid Dynamics}

\begin{document}

\title{Exact instantaneous optimals in the non-geostrophic Eady problem and the detrimental effects of discretization\thanks{WB is supported by the US National Science Foundation grant DMS 1407340.}
}

\titlerunning{Eady Instantaneous Optimals}        

\author{William Barham         \and
        Ian Grooms
}


\institute{W.~Barham and I.~Grooms \at
              Department of Applied Mathematics\\
              University of Colorado, Boulder, CO, 80309, USA\\
              Tel: +1303-492-7563\\
              \email{ian.grooms@colorado.edu}
}

\date{Received: date / Accepted: date}

\maketitle
\begin{abstract}
We derive exact analytical expressions for flow configurations that optimize the instantaneous growth rate of energy in the linear Eady problem, along with the associated growth rates.
These optimal perturbations are relevant linear stability analysis, but, more importantly, they are relevant for understanding the energetics of fully nonlinear baroclinic turbulence.
The optimal perturbations and their growth rates are independent of the Richardson number.
The growth rates of the optimal perturbations grow linearly as the horizontal wavelength of the perturbation decreases.
Perturbation energy growth at large scales is driven by extraction of potential energy from the mean flow, while at small scales it is driven by extraction of kinetic energy from the mean shear.
We also analyze the effect of spatial discretization on the optimal perturbations and their growth rates.
A second order energy-conserving discretization on the Arakawa B grid generally has too-weak growth rates at small scales and is less accurate than two second order discretizations on the Arakawa C grid.
The two C grid discretizations, one that conserves energy and another that conserves both energy and enstrophy, yield very similar optimal perturbation growth rates that are significantly more accurate than the B grid discretization at small scales.
\end{abstract}

\section{Introduction}
\label{intro}
The Eady problem is one of the canonical hydrodynamic stability problems in geophysical fluid dynamics \cite{Eady49,Vallis17}.
The equilibrium solution whose stability is examined consists of a Boussinesq fluid in a horizontal plane layer with vertical rotation and gravity, with a horizontally-uniform velocity with constant vertical shear.
The Coriolis acceleration induced by this velocity is exactly balanced by a horizontal pressure gradient where the pressure is in hydrostatic balance with a density profile that varies linearly in both the horizontal and vertical directions.
The linear stability problem is analytically solvable in the quasigeostrophic approximation, and the classical result is that the equilibrium is unstable to infinitesimal perturbations whose exponential growth rate scales with the amplitude of the background flow, i.e.~the equilibrium is unstable unless the fluid is at rest.
This instability that appears in the quasigeostrophic Eady problem is an example of baroclinic instability.
Stone \cite{Stone66,Stone70} relaxed the quasigeostrophic approximation and found modifications of the baroclinic instability at low Richardson number (i.e.~strong shear in comparison with the stratification) and a `symmetric' instability that is completely absent in the quasigeostrophic approximation and that only occurs at low Richardson numbers.

As frequently occurs in hydrodynamic linear stability problems, the linear evolution operator in the Eady problem is non-normal, i.e.~it does not commute with its adjoint with respect to the $L^2$ inner product.
The solutions to linear initial value problems with non-normal operators can exhibit growth even when the operator has no eigenvalues with positive real part \cite{TE05}, and non-normal growth has been invoked in some scenarios to explain subcritical transition to turbulence \cite{BB88,SH01}.
Schmid \cite{Schmid07} reviews linear hydrodynamic stability in the presence of non-normality.
Non-normality seems at first glance to be irrelevant to the Eady problem since the latter is always unstable to exponentially-growing, i.e.~normal-mode, perturbations.
Nevertheless, Farrell \cite{Farrell84,Farrell85} demonstrated that non-normality in the (damped) Eady problem is more important than exponential behavior in the short-term growth of perturbations in the linear problem.\\

The most energetic eddies in the ocean are thought to obtain most of their energy by extracting potential energy from the large-scale flow in a manner similar to the extraction of potential energy from the equilibrium state by unstable perturbations in linear baroclinic instability \cite{FW09}.
Many studies (e.g.~\cite{Smith07,TMHS11}) catalogue the properties of linear quasigeostrophic baroclinic instability across the world oceans in hopes of learning something about the properties of the strongly nonlinear ocean eddies.
But there is no immediately obvious mathematical connection between fully nonlinear turbulence and exponential or non-normal growth of infinitestimal perturbations about an equilibrium solution.
A direct mathematical connection between the linear problem and fully nonlinear statistically stationary turbulence was nevertheless provided by DelSole \cite{DelSole04}, which we here recall.

Consider an autonomous, unforced fluid system written schematically as
\begin{gather}
\pd{t}\bu = \mathcal{L}\bu + \mathcal{B}(\bu,\bu)
\end{gather}
where $\mathcal{L}$ includes the linear terms like viscous dissipation and $\mathcal{B}(\bu,\bu)$ is a bilinear operator that conserves energy.
The energy is assumed to be in the form
\begin{equation}
E = \frac{1}{2}\langle \bu,\bu\rangle
\end{equation}
for some inner product $\langle\bcdot,\bcdot\rangle$.
Now let $\bar{\bu}$ be a steady exact solution of the governing equations and consider the linear evolution of infinitestimal perturbations $\bu'$.
The perturbations $\bu'$ evolve according to 
\begin{gather}
\pd{t}\bu' = \mathcal{L}\bu' + \mathcal{B}(\bar{\bu},\bu')+\mathcal{B}(\bu',\bar{\bu})  = \overline{\mathcal{L}}\bu'.
\end{gather}
The energy in the perturbations evolves according to
\begin{multline}
\dd{}{t}{}E = \frac{1}{2}\langle\bu',\pd{t}\bu'\rangle + \frac{1}{2}\langle\pd{t}\bu',\bu\rangle = \frac{1}{2}\langle\bu',\overline{\mathcal{L}}\bu'\rangle + \frac{1}{2}\langle\overline{\mathcal{L}}\bu',\bu'\rangle \\
= \left\langle\bu',\frac{\overline{\mathcal{L}}^\dag+\overline{\mathcal{L}}}{2}\bu'\right\rangle\label{eqn:IntroEnergy}
\end{multline}
where the symbol $\overline{\mathcal{L}}^\dag$ denotes the adjoint of $\mathcal{L}$ with respect to the energy inner product.
The operator $(\overline{\mathcal{L}}^\dag+\overline{\mathcal{L}})/2$ is clearly self-adjoint even when $\overline{\mathcal{L}}$ is non-normal, so all of its eigenvalues must be real and the eigenfunctions associated with distinct eigenvalues must be orthogonal.
The eigenfunction corresponding to the most-positive eigenvalue of $(\overline{\mathcal{L}}^\dag+\overline{\mathcal{L}})/2$ is called the `instantaneous optimal' and maximizes the instantaneous growth rate of energy over all perturbations with the same amplitude.

Now suppose that we repeat the above analysis but instead interpret $\bar{\bu}$ as the time-average in statistically stationary but fully nonlinear turbulence, and $\bu'=\bu-\bar{\bu}$ as the perturbations about the time average.
These perturbations evolve according to
\begin{equation}
\pd{t}\bu' = \mathcal{L}\bu' + \mathcal{B}(\bar{\bu},\bu')+\mathcal{B}(\bu',\bar{\bu}) + \mathcal{B}(\bu',\bu')' = \overline{\mathcal{L}}\bu' + \mathcal{B}(\bu',\bu')'.
\end{equation}
The eddy-eddy nonlinearity $\mathcal{B}(\bu',\bu')'$ conserves energy, so the energy equation for the perturbations is exactly the same as before, i.e.~equation (\ref{eqn:IntroEnergy}).
If the operator $(\overline{\mathcal{L}}^\dag+\overline{\mathcal{L}})/2$ is negative definite, then the energy must decay and cannot remain statistically steady; one expects statistically-steady turbulence to correspond to a setting where the operator has at least one positive eigenvalue.
This connection between the linear theory of instantaneous optimals and statistically stationary fully nonlinear turbulence was first made by DelSole \cite{DelSole04}.
It enables research into the energetics of fully-developed turbulence using analysis of linear, self-adjoint operators.

It bears noting that it is not always trivial to compute the operator $\overline{\mathcal{L}}^\dag$.
The point spectrum (i.e.~eigenvalues and eigenfunctions) of the operator $\overline{\mathcal{L}}^\dag+\overline{\mathcal{L}}$ is related to the following constrained optimization problem: Optimize the instantaneous energy growth rate $\langle\bu',(\overline{\mathcal{L}}^\dag+\overline{\mathcal{L}})\bu'\rangle$ subject to the constraint $\langle\bu',\bu'\rangle=1$.
The Euler-Lagrange equations for this optimization problem are
\begin{equation}
(\overline{\mathcal{L}}^\dag+\overline{\mathcal{L}})\bu'=\lambda\bu'
\end{equation}
which makes clear the connection to the point spectrum of $\overline{\mathcal{L}}^\dag+\overline{\mathcal{L}}$, but the Euler-Lagrange equations can sometimes be derived without explicitly finding an expression for $\overline{\mathcal{L}}^\dag$.
This approach is used in section \ref{sec:Exact}.\\

In this paper we begin with a derivation of exact analytical expressions for the instantaneous optimals of the non-geostrophic Eady problem.
Instantaneous optimals for the quasigeostrophic Eady problem have been computed by Farrell and Ioannou \cite{Farrell89,FI96} by first discretizing the quasigeostrophic linear problem and then computing instantaneous optimals of the discretized system.
The quasigeostrophic Eady problem was recently analyzed directly by Kalashnik and Chkhetiani \cite{KC18}.
Our results are similar to some extent, but we uncover a new class of instantaneous optimals that is absent from the quasigeostrophic problem.

We continue the study by computing instantaneous optimals for the spatially-discrete problem using three discretizations commonly used in global ocean models.
The analysis extends the authors' previous work on the discrete linear stability problem \cite{BBG18}, which studied only exponentially-growing solutions.
The impact of spatial discretization on exponentially-growing linear instability has also been explored in geophysical scenarios in \cite{AM88,BW88,BW17,DlSMB17,HKRB83}.

The exact analytical expressions for instantaneous optimals in the non-geostrophic Eady problem are derived in section \ref{sec:Exact}.
Instantaneous optimals in a discrete version of the problem are then analyzed in section \ref{sec:Discrete}.
The results are discussed and conclusions are offered in section \ref{sec:Conclusions}.

\section{Analytical Instantaneous Optimals}
\label{sec:Exact}
The nondimensional hydrostatic non-geostrophic linear perturbation equations in the Eady problem are \cite{Grooms15}
\begin{align}\label{eqn:A1}
\left(\pd{t}+z\pd{x}\right)\bu' +\eps w'\hat{\bm{x}}+ \eps^{-1}(\hat{\bm{z}}\times\bu')_h &= -\eps^{-1}\bnabla_h p'\\\label{eqn:A2}
\pd{z}p'&=b'\\
\left(\pd{t}+ z\pd{x}\right)b' - v' + w' &= 0\\
\bnabla\bcdot\bu' + \eps\pd{z}w'&=0\label{eqn:A3}
\end{align}
where the Brunt-V\"ais\"al\"a frequency is $N$, the dimensional background velocity is $\bar{u} = \Lambda z$, the Richardson number is $\eps^{-2}$, and $\eps=\Lambda/N$.
Time is nondimensionalized via the time scale $N/(f\Lambda)$ where $f$ is the Coriolis parameter, and horizontal directions are nondimensionalized via the deformation radius $NH/f$ where $H$ is the depth of the fluid.
The vertical velocity $w'$ has been scaled to be a factor of $f\Lambda/N^2$ smaller than the horizontal velocity $\bu'$.
The subscript $_h$ denotes the horizontal component of a vector, e.g.~$\bnabla_h = (\pd{x},\pd{y})$.

The perturbation energy equation is obtained by taking the dot product of (\ref{eqn:A1}) with $\bu'=(u',v')$, multiplying (\ref{eqn:A2}) by $b'$, adding, and integrating over the domain $\Omega$
\begin{equation}
\frac{1}{2}
\dd{}{t}{}\int_\Omega(u')^2+(v')^2 +(b')^2 = \int_\Omega v'b'-\eps w'u'
\end{equation}
where $\int_\Omega$ denotes an integral over the physical domain $\Omega$.
The goal is to obtain a configuration of $(u',v',b')$ that maximizes the growth rate of the energy over all configurations with unit energy.
We therefore define the Lagrangian
\begin{equation}
I[u', v', b'] = B+S-\lambda E
\end{equation}
where $\lambda$ is the Lagrange multiplier and the growth rate has been split into two components
\begin{equation}
B = \int_\Omega v'b',\;\;S = -\eps\int_\Omega w'u'
\end{equation}
where $B$ is the baroclinic conversion and $S$ is shear production.
As usual, the Euler-Lagrange equations are derived by finding stationary points of the Lagrangian.

The Euler-Lagrange equations for this constrained optimization problem are obtained as follows.
We first consider the Frechet derivatives of the energy, $E$, and of the baroclinic conversion $B$, which are simply
\begin{align}
\text{d}E = \int_\Omega u'\delta u+v'\delta v+b'\delta b,\text{ and }\text{d}B = \int_\Omega v'\delta b + b'\delta v.
\end{align}
The derivative of the shear production, $S$, is less straightforward, since the vertical velocity $w'$ is not independent of the horizontal velocity but is instead derived from
\begin{equation}
w' = -\eps^{-1}\int_0^z\pd{x}u'+\pd{y}v'\text{d}s.
\end{equation}
To derive the Frechet derivative of the shear production, we need the following simple integration by parts identity
\begin{equation}
\int_0^1g(z)\int_0^zh(s)\text{d}s\text{d}z = \left(\int_0^1g(z)\text{d}z\right)\left(\int_0^1h(z)\text{d}z\right)-\int_0^1h(z)\int_0^zg(s)\text{d}s\text{d}z
\end{equation}
which is valid for integrable functions $g$ and $h$.
The $s$ variable is a stand-in for the vertical coordinate.
With this expression in hand, note that
\begin{align}
\eps\int_0^1 u'\delta w \text{d}z&= -\int_0^1u'\int_0^z\left(\pd{x}\delta u+\pd{y}\delta v\right)\text{d}s\text{d}z\\
&=-\left(\int_0^1u'\text{d}z\right)\left(\int_0^1\left(\pd{x}\delta u+\pd{y}\delta v\right)\text{d}z\right) + \int_0^1\left(\pd{x}\delta u+\pd{y}\delta v\right)\int_0^zu'\text{d}s\text{d}z.
\end{align}
The fact that $\delta w=0$ at both the upper and lower boundaries sets the boundary term to zero.
The Frechet derivative of the shear production is therefore
\begin{equation}
\text{d}S =-\int_\Omega \left( \int_0^z2\pd{x}u'+\pd{y}v'\text{d}s\right)\delta u + \left(\pd{y}u'\right)\delta v.
\end{equation}
Configurations of $(u',v',b')$ that are stationary with respect to the energy growth rate subject to the condition of unit energy satisfy
\begin{equation}
\text{d}I = \text{d}B+\text{d}S - \lambda\text{d}E = 0
\end{equation}
for all $(\delta u, \delta v,\delta b)$.
The Euler-Lagrange equations are therefore
\begin{gather}
\int_0^z\left(2\pd{x}u'+\pd{y}v'\right) = \lambda u'\\
b'+\int_0^z\pd{y}u' = \lambda v'\\
v'=\lambda b'.
\end{gather}
Perturbation fields $(u',v',b')$ that satisfy these equations are associated with energy growth rates $\lambda$.
In the following we make use of the following equations, obtained by taking a single derivative with respect to $z$
\begin{gather}\label{eqn:EL1}
2\pd{x}u'+\pd{y}v' = \lambda \pd{z}u'\\\label{eqn:EL2}
\pd{z}b'+\pd{y}u' = \lambda \pd{z}v'\\\label{eqn:EL3}
v'=\lambda b'.
\end{gather}

\subsection{Perturbations on  the baroclinic axis}
This section considers perturbations with no dependence on the $y$ coordinate.
Perturbations of this type satisfy $v'=\lambda b'$ and $b=\lambda v'$ which requires either $\lambda=1$ and $v'=b'$ or $v'=b'=0$.

Perturbations with $v'=b'=0$ must satisfy $2\pd{x}u'-\lambda\pd{z}u'=0$.
The general solution for these perturbations has the form
\begin{equation}
u'=\cos\left(k_x x +\frac{2k_x}{\lambda}z + \phi\right) 
\end{equation}
where $k_x$ is any real wavenumber and $\phi$ is any real phase shift.
The vertical velocity associated with perturbations of this form must have the form
\begin{equation}
w' = \frac{\lambda}{2\eps}\left[\cos\left(k_x x+\phi\right) - \cos\left(k_x x+\frac{2k_x}{\lambda}z + \phi\right)\right].
\end{equation}
The condition that $w'=0$ at $z=1$ requires
\begin{equation}
\lambda = \frac{|k_x|}{\pi m}
\end{equation}
for any nonzero integer $m$.
The maximal growth rate is obtained for $m =$ sign$(k_x)$.\\

In summary, optimal perturbations with no $y$-dependence come in two forms.
One form has $v'=b'$ and $u'=w'=0$, with growth rate $\lambda=1$ regardless of the horizontal scale or vertical structure of the perturbations; it derives its energy solely from baroclinic energy conversion.
The other form has $v'=b'=0$ and derives its energy solely from shear production; its optimal growth rate increases as the horizontal scale decreases.
Overall, the fastest-growing perturbation is the baroclinic form for $|k_x|\le\pi$ and has growth rate $\lambda=1$; for $|k_x|>\pi$ the ageostrophic mode dominates with a growth rate $\lambda = |k_x|/\pi$.

\subsection{Perturbations on the symmetric axis}
\label{sec:Symmetric}
This section considers perturbations with no dependence on the $x$ coordinate.
Perturbations of this form must satisfy
\begin{equation}
\pd{y}^2u' + (1-\lambda^2)\pd{z}^2u'=0.
\end{equation}
The general solution now takes the form
\begin{equation}
u' = \alpha\cos\left(k_y y +\frac{k_y}{\sqrt{\lambda^2-1}}z+\phi_1\right) + \beta\cos\left(k_y y -\frac{k_y}{\sqrt{\lambda^2-1}}z+\phi_2\right)
\end{equation}
for arbitrary $\alpha$ and $\beta$ and phase shifts $\phi_1$ and $\phi_2$.
The other components of the solution are obtained from the Euler-Lagrange equations and take the form
\begin{gather}
v' = \frac{\alpha\lambda}{\sqrt{\lambda^2-1}}\cos\left(k_y y +\frac{k_y}{\sqrt{\lambda^2-1}}z+\phi_1\right)- \frac{\beta\lambda}{\sqrt{\lambda^2-1}}\cos\left(k_y y -\frac{k_y}{\sqrt{\lambda^2-1}}z+\phi_2\right)\\
b'=\frac{v'}{\lambda}
\end{gather}
\begin{multline}
w'=\frac{\alpha\lambda}{\eps}\left[\cos(k_yy+\phi_1)-\cos\left(k_y y +\frac{k_y}{\sqrt{\lambda^2-1}}z+\phi_1\right)\right]\\
+\frac{\beta\lambda}{\eps}\left[\cos(k_yy+\phi_2)-\cos\left(k_y y -\frac{k_y}{\sqrt{\lambda^2-1}}z+\phi_2\right)\right].
\end{multline}
There are now two ways to impose $w'=0$ at $z=1$.
The simplest method is require $k_y/\sqrt{\lambda^2-1} = 2\pi m$ for any integer $m$; this sets both components of the solution to 0 at the boundary.
The other method requires coordination between the two components of the solution; specifically, the condition can be achieved by taking $\phi_1=\phi_2$, $\beta=-\alpha$, and requiring the arguments of the cosines to differ by $2\pi m$ for any integer $m$.
This condition yields
\begin{equation}
\lambda = \pm\left(\frac{k_y^2}{\pi^2m^2}+1\right)^{1/2}.
\end{equation}
The simpler method of imposing $w'=0$ at $z=1$ mentioned above merely yields a subset of the growth rates associated with the more general solution.
The growth rate is clearly maximized at $m=\pm1$ and by taking the positive sign of $\lambda$.\\

The baroclinic and shear-production components of the energy growth rate associated with these perturbations are
\begin{equation}
B = \alpha^2\left(\frac{\pi(k_y^2+\pi^2)^{1/2}}{k_y^2}\right),\quad S = \alpha^2\left(\frac{(k_y^2+\pi^2)^{1/2}}{\pi}\right).
\end{equation}
The ratio of the different components of the energy growth rate is
\begin{equation}
\frac{B}{S} = \frac{\pi^2}{k_y^2}.
\end{equation}
The baroclinic production is larger than the shear production for $|k_y|<\pi$, and vice versa for $|k_y|>\pi$.\\

In summary, for optimal perturbations with no $x$-dependence, the growth rate is always larger than 1 and behaves as $|k_y|\pi$ in the limit $k_y\to\infty$.
Both baroclinic and shear production contribute to the energy growth at all scales, but baroclinic production dominates for optimal perturbations with $k_y<\pi$ while shear production dominates for optimal perturbations with $k_y>\pi$.
The optimal perturbations have sinusoidal vertical structure with a half of a wavelength fitting into the total fluid depth.

\subsection{General perturbations}
Perturbations with both $x$ and $y$ dependence must satisfy
\begin{equation}
\left[\lambda\pd{y}^2+2(\lambda^2-1)\pd{x}\pd{z}+\lambda(1-\lambda^2)\pd{z}^2\right]b'=0.
\end{equation}
A general solution can be written as the sum of the following two linearly-independent basic solutions
\begin{align}
b_1 &= \cos\left(k_x x + k_y y + \left(\frac{k_x}{\lambda} + \left(\frac{k_x^2}{\lambda}+\frac{k_y^2}{\lambda^2-1}\right)^{1/2}\right)z + \phi_1\right)\\
b_2 &= \cos\left(k_x x + k_y y + \left(\frac{k_x}{\lambda} - \left(\frac{k_x^2}{\lambda}+\frac{k_y^2}{\lambda^2-1}\right)^{1/2}\right)z + \phi_2\right).
\end{align}
The remaining components of the solution can be derived from $b'$ using the Euler-Lagrange equations (\ref{eqn:EL1})--(\ref{eqn:EL3}).
Analytical formulas are easily available, but are sufficiently cumbersome that they are here omitted, with the exception of the vertical velocity at $z=1$
\begin{equation}
w_1 =-\frac{k_y^2\lambda+k_x\gamma_+(\lambda^2-1)}{k_y\eps\gamma_+}\left[\cos(k_xx+k_yy+\phi_1)-\cos(k_xx+k_yy+\phi_1+\gamma_+z)\right]
\end{equation}
\begin{multline}
w_2 =-\frac{k_y^2\lambda+k_x\gamma_-(\lambda^2-1)}{k_y\eps\gamma_-}\left[\cos(k_xx+k_yy+\phi_2)-\cos(k_xx+k_yy+\phi_2+\gamma_-z)\right]
\end{multline}
where
\begin{equation}
\gamma_\pm = \frac{k_x}{\lambda}\pm\left(\frac{k_x^2}{\lambda^2}+\frac{k_y^2}{\lambda^2-1}\right)^{1/2}.
\end{equation}

As in the previous subsection, there are two ways of imposing the condition that $w=0$ at $z=1$.
The simplest method is to require $\gamma_+=2\pi m$, which sets $w_1=0$ at $z=1$, and to require the solution to include only $b_1'$, $u_1'$, and $v_1'$; a similar solution is available by setting $\gamma_-=2\pi m$.
As in the previous section there is another way to set $w'=0$ at $z=1$ that requires coordination among the two components of the solution.
Specifically, the condition can be achieved by taking $\phi_1=\phi_2$ and then choosing a linear combination of $w_1$ and $w_2$ such that the terms $\cos(k_xx+k_yy+\phi_1)$ and $\cos(k_xx+k_yy+\phi_2)$ cancel.
The remainder is then proportional to $\cos(k_xx+k_yy+\phi_1+\gamma_+)-\cos(k_xx+k_yy+\phi_2+\gamma_-)$, which is zero provided that $\gamma_+-\gamma_-=2\pi m$ for any integer $m$.
This leads to the condition
\begin{equation}
\left(\frac{k_x^2}{\lambda^2}+\frac{k_y^2}{\lambda^2-1}\right)^{1/2}=\pi m.
\end{equation}
The solutions have the form
\begin{equation}
\lambda^2 = \frac{k_x^2+k_y^2+m^2\pi^2\pm\left((k_x^2+k_y^2+m^2\pi^2)^2-4k_x^2m^2\pi^2\right)^{1/2}}{2m^2\pi^2}.
\end{equation}
Solutions are decreasing functions of $m$ so the maximal growth rate is at $m=\pm1$, and has the form
\begin{equation}
\lambda = \left(\frac{k_x^2+k_y^2+\pi^2+\left((k_x^2+k_y^2+\pi^2)^2-4k_x^2\pi^2\right)^{1/2}}{2\pi^2}\right)^{1/2}.\label{eqn:General}
\end{equation}
In the limit $k_x\sim k_y\gg1$, the behavior is $\lambda\sim (k_x^2+k_y^2)^{1/2}/\pi$.\\

\begin{figure*}
  \includegraphics[width=\textwidth]{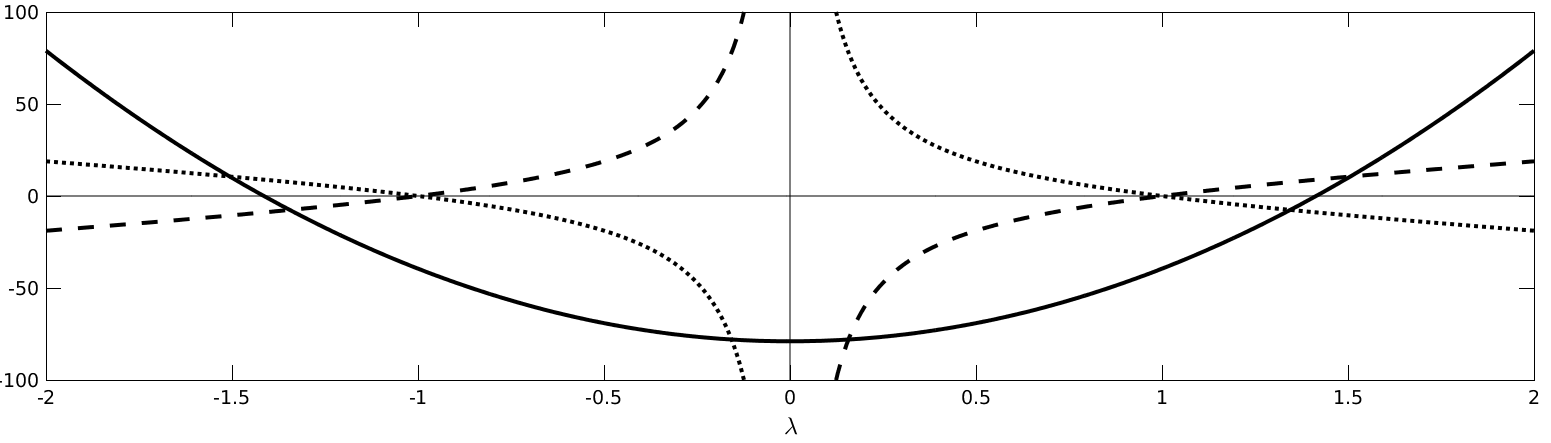}
\caption{The function on the right of (\ref{eqn:lambda1}) is shown as a solid line; the function on the left of (\ref{eqn:lambda1}) with $k_xm>0$ is shown as a dashed line; the function on the left of (\ref{eqn:lambda1}) with $k_xm<0$ is shown as a dotted line. All lines use $k_y=2\pi$ and $k_x=1$. Eigenvalues $\lambda$ occur at the intersections of the lines.}
\label{fig:lambda1}       
\end{figure*}

We now analyze the other way of imposing the condition $w'=0$ at $z=1$, i.e. by setting $\gamma_+=2\pi m$ or $\gamma_-=2\pi m$.
In both cases the eigenvalue $\lambda$ must be a root of the following cubic
\begin{equation}
4m^2\pi^2\lambda^3-4k_xm\pi\lambda^2-(4m^2\pi^2+k_y^2)\lambda+4 k_x m\pi=0.\label{eqn:lambda0}
\end{equation}
The number and locations of the real roots of this cubic are not immediately apparent, so the condition is rephrased as the value of $\lambda$ at the intersection of two curves
\begin{equation}
4\pi k_xm\left(\frac{\lambda^2-1}{\lambda}\right) = 4 \pi^2m^2(\lambda^2-1)-k_y^2.\label{eqn:lambda1}
\end{equation}
The curve corresponding to the right hand side of this expression is simply a parabola opening upwards while the curve corresponding to the expression on the left is a hyperbola whose primary axes are tilted.
An example of these curves with $k_y=2\pi$, $k_x=1$, and $m=\pm1$ is shown in the figure \ref{fig:lambda1}.
The function $4\pi k_xm(\lambda^2-1)\lambda^{-1}$ has roots at $\lambda=\pm1$, while the function $4 \pi^2m^2(\lambda^2-1)-k_y^2$  has roots at $\lambda = \pm \sqrt{1+k_y^2/(4\pi^2m^2)}$.
The curves always intersect at three values of $\lambda$, which implies that the cubic (\ref{eqn:lambda0}) always has three real roots.
If $k_xm>0$, there are two positive intersections and one negative intersection; if $k_xm<0$, the signs are reversed.
If $k_xm>0$, the negative intersection occurs to the right of the root of the parabola at $\lambda = -\sqrt{1+k_y^2/(4\pi^2m^2)}$; the smaller positive intersection occurs to the left of the root of the hyperbola at $\lambda=1$; and the larger positive intersection  occurs to the right of the root of the parabola at $\lambda = \sqrt{1+k_y^2/(4\pi^2m^2)}$.
The largest root, therefore, occurs when $k_x$ and $m$ have the same sign, and we have the following lower bound on the growth rate of the optimal perturbation
\begin{equation}
\lambda_\text{max}\ge \sqrt{1+k_y^2/(4\pi^2m^2)}.\label{eqn:lambdaBound}
\end{equation}

Clearly the lower bound is decreasing in $m$ so that the lower bound is optimized for $m=1$.
To obtain a firmer understanding of how the actual growth rate depends on $m$, the cubic equation for $\lambda$ is re-written as follows
\begin{equation}
\frac{k_y^2}{4\pi^2m^2} + \frac{k_x}{\pi m}\left(\frac{\lambda^2-1}{\lambda}\right) = \lambda^2-1.\label{eqn:lambda2}
\end{equation}
The function on the right is simply an upwards-opening parabola.
The function on the left is a hyperbola whose axes are tilted and that is shifted downwards.
Increasing $m$ causes every point on the hyperbola to move downwards, which makes every point of intersection between the hyperbola and the parabola move towards $\lambda=0$.
The actual optimal growth rate is therefore a decreasing function of $m$, and the optimal growth rate at any value of $k_x$ and $k_y$ is obtained by setting $m =$ sign$(k_x)$.\\

We have analyzed two ways of obtaining optimal perturbations and their associated growth rates.
The first class of solutions analyzed leads to growth rates $\lambda$ scaling as $\lambda\sim (k_x^2+k_y^2)^{1/2}/\pi$ in the limit $k_x\sim k_y\gg1$.
For the second class of solutions, the dominant balance in (\ref{eqn:lambda0}) leads to an optimal growth rate of the form $\lambda\sim (|k_x|+\sqrt{k_x^2+k_y^2})/(2\pi)$, which is smaller than the first method.
Similarly, taking the limit $k_x\to0$ in the second method yields a lower growth rate than the optimal rate computed in section \ref{sec:Symmetric}.
We conclude that the first method yields the optimal solution with growth rates given by (\ref{eqn:General}).


\section{Discrete Instantaneous Optimals}
\label{sec:Discrete}
This section follows the authors' previous work \cite{BBG18} in studying the impact of spatial discretization on the linear non-geostrophic Eady problem.
The previous work studied only exponentially-growing modes; as discussed in the introduction, instantaneous optimals are more relevant to the dynamics of fully nonlinear turbulence.
In the next subsection we recall the linear perturbation equations associated with the three discretizations from \cite{BBG18}.
The following subsection then proceeds to compute instantaneous optimals using the three discretizations at various resolutions.

\subsection{Spatial Discretizations}
\label{sec:BBG18}
In \cite{BBG18} the authors considered several spatial discretizations that are commonly used in $z$-coordinate global ocean models.
All the discretizations considered here and in \cite{BBG18} use the same discretization of the vertical coordinate $z$; it is the standard second-order discretization \cite{Bryan69,HB99} of the $z$ direction used in $z$-coordinate ocean models (e.g.~the Modular Ocean Model \cite{Griffies04}, the Parallel Ocean Program \cite{POP}, and the Nucleus for European Modeling of the Ocean \cite{Madec08}).
The variables $u'$, $v'$, $b'$, and $p'$ are all discretized on an equispaced grid in $z$ with $N_z$ points, where the values on the grid are contained in the vectors $\bm{U}$, $\bm{V}$, $\bm{b}$, and $\bm{p}$, respectively (the grid is equispaced here for simplicity; global ocean models typically use uneven spacing).
The values $\bm{U}_k$ for $k=1,\ldots,N_z$ correspond to the values of $u'$ at the levels $z_k=(k-1/2)/N_z$.
The depth-averaged component of pressure is treated separately from the baroclinic component of pressure, so the total discrete pressure at level $z_k$ is denoted $\bm{p}_k+\phi$.

The baroclinic part of the pressure $p'$ is determined from the hydrostatic balance $\pd{z}p'=b'$.
The hydrostatic balance is discretized by making a centered finite-difference approximation to $\pd{z}p'$ at the levels $z_k'=k/N_z$ equal to the value of $b'$ linearly interpolated to the levels $z_k'$, i.e.
\begin{equation}
N_z(\bm{p}_{k+1}-\bm{p}_k) = \frac{\bm{b}_{k+1}+\bm{b}_k}{2},\;\;k=1,\ldots,N_z-1.
\end{equation}
There are only $N_z-1$ equations here, which are insufficient to determine the pressure at each of the $N_z$ levels; the remaining condition is that the baroclinic pressure should integrate to zero across the depth, which is enforced via a simple midpoint-rule quadrature, i.e.
\begin{equation}
\sum_{k=1}^{N_z}\bm{p}_k=0.
\end{equation}
The complete set of equations for baroclinic pressure thus takes the form
\begin{equation}
N_z\left[\begin{array}{ccccc}
-1&1&0&\cdots&0\\
0&\ddots&\ddots&&\vdots\\
\vdots&&\ddots&\ddots&\vdots\\
0&\cdots&0&-1&1\\
1&\cdots&\cdots&\cdots&1
\end{array}\right]\bm{p} = \frac{1}{2}\left[\begin{array}{ccccc}
1&1&0&\cdots&0\\
0&\ddots&\ddots&&\vdots\\
\vdots&&\ddots&\ddots&\vdots\\
0&\cdots&0&1&1\\
0&\cdots&\cdots&\cdots&0
\end{array}\right]\bm{b}.
\end{equation}
It is convenient to write this system as $\bm{p} = \mat{P}\bm{b}$, where 
\begin{equation}
\mat{P} = \frac{1}{2N_z}\left[\begin{array}{ccccc}
-1&1&0&\cdots&0\\
0&\ddots&\ddots&&\vdots\\
\vdots&&\ddots&\ddots&\vdots\\
0&\cdots&0&-1&1\\
1&\cdots&\cdots&\cdots&1
\end{array}\right]^{-1}\left[\begin{array}{ccccc}
1&1&0&\cdots&0\\
0&\ddots&\ddots&&\vdots\\
\vdots&&\ddots&\ddots&\vdots\\
0&\cdots&0&1&1\\
0&\cdots&\cdots&\cdots&0
\end{array}\right].
\end{equation}

The value of vertical velocity $w'$ is tracked in the models at levels $z_k' = k/N_z$ that are staggered with respect to the levels $z_k$ where the other variables are tracked.
However, the values of $w'$ that appear in the linear perturbation equations are the values at $z_k'$ linearly interpolated to $z_k$ (for a complete discussion see \cite{BBG18}).
The values of $w'$ that appear in the linear perturbation equations are obtained by applying the following second-order quadrature to $w'(z) = -\eps^{-1}\int_0^z\pd{x}u'+\pd{y}v'$
\begin{equation}
w'(z_k)\approx -\frac{1}{N_z}\sum_{j=1}^k\left[\pd{x}\bm{U}_j + \pd{y}\bm{V}_j\right] - \frac{1}{2N_z}\left(\pd{x}\bm{U}_k + \pd{y}\bm{V}_k\right).
\end{equation}
Defining $\bm{W}$ to be the vector containing these approximate values of $w'(z_k)$, we may write the above expression as a matrix-vector multiplication
\begin{equation}
\bm{W} = -\frac{1}{2\eps N_z}\left[\begin{array}{ccccc}
1&0&\cdots&\cdots&0\\
2&1&0&&\vdots\\
\vdots&\ddots&\ddots&\ddots&\vdots\\
2&\cdots&2&1&0\\
2&\cdots&\cdots&2&1\end{array}\right]\left(\pd{x}\bm{U}+\pd{y}\bm{V}\right) = -\frac{1}{\eps}\mat{W}\left(\pd{x}\bm{U}+\pd{y}\bm{V}\right).
\end{equation}

The vertically-discretized linear perturbation equations for can now be written as
\begin{align}\label{eqn:D1}
\pd{t}\bm{U}&= -\bm{Z}\pd{x}\bm{U}+\mat{W}(\pd{x}\bm{U}+\pd{y}\bm{V})+\eps^{-1}\bm{V}-\eps^{-1}\mat{P}\pd{x}\bm{b}-\eps^{-1}\bm{1}\pd{x}\phi\\\label{eqn:D2}
\pd{t}\bm{V}&= -\bm{Z}\pd{x}\bm{V}-\eps^{-1}\bm{U}-\eps^{-1}\mat{P}\pd{y}\bm{b}-\eps^{-1}\bm{1}\pd{y}\phi\\\label{eqn:D3}
\pd{t}\bm{b}&= -\bm{Z}\pd{x}\bm{b}+\bm{V}+\eps^{-1}\mat{W}(\pd{x}\bm{U}+\pd{y}\bm{V})
\end{align}
where $\mat{Z}$ is a diagonal matrix with entries $z_k$.
The discrete barotropic pressure is $\phi\bm{1}$ where $\bm{1}$ is a vector with every entry equal to one.
The discrete barotropic pressure is set by the condition that the divergent component of the depth-integrated horizontal velocity should not evolve, i.e.
\begin{equation}
\nabla_h^2\phi = \frac{1}{N_z}\bm{1}\bcdot\left[\eps\left(\mat{W}-\mat{Z}\right)\pd{x}(\pd{x}\bm{U}+\pd{y}\bm{V})+\pd{x}\bm{V}-\pd{y}\bm{U}\right].\label{eqn:D4}
\end{equation}
This equation is derived by taking the dot product of $\bm{1}/N_z$ with (\ref{eqn:D1}) and (\ref{eqn:D2}) and relies on the facts that taking the dot product of a vector with $\bm{1}/N_z$ is equivalent to a midpoint-rule approximation to the integral across depth, that $\bm{1}\bcdot\bm{1}=N_z$, and that $\bm{1}\bcdot\mat{P}\bm{b}=0$ because $\mat{P}\bm{b}$ is the baroclinic pressure.

It should be noted that the barotropic pressure does not prevent us from specifying an initial condition for the discrete dynamics that has a divergent depth-integrated component; it simply guarantees that that component will not evolve.
In an ocean model this can be dealt with by the simple expedient of specifying an initial condition that is fully incompressible; similarly, in our setting, it will be required that initial perturbations with divergent barotropic component be precluded.
This issue was avoided in \cite{BBG18} because that paper only studied exponentially-growing solutions, and the barotropic pressure already prevents a divergent barotropic component from evolving.
More care is required here since an initial perturbation with a divergent barotropic component can impact the initial energy growth rate even if the divergent barotropic component does not evolve itself.\\

To complete the analysis, the linear perturbation equations (\ref{eqn:D1})--(\ref{eqn:D4}) are further discretized in the horizontal directions.
We consider the same three horizontal discretizations used in \cite{BBG18}.
The first uses the Arakawa B grid \cite{AL77}, where the horizontal velocity is discretized at locations half of a grid length north and east of the buoyancy.
The B grid discretization is second-order and energy-conserving and is due to Bryan \cite{Bryan69}.
The remaining two discretizations use the Arakawa C grid, where $u'$ is discretized half of a grid length east of the buoyancy and $v'$ is discretized half a grid point north of the buoyancy.
The second method is the C grid equivalent of the B grid method from \cite{Bryan69}; it is second-order and conserves energy.
The third method is a second-order energy and enstrophy conserving discretization based on the vector-invariant formulation of the momentum equations \cite{AL81,lSPTMB09}; it will be referred to as the EEN scheme (Energy and ENstrophy) following \cite{BBG18,lSPTMB09}.

In a domain of width $L$ with an equispaced square grid of size $\Delta_x$, the discrete wavenumbers from $2\pi/L$ to $\pi/\Delta_x$ are represented.
Taking the domain to be of infinite extent allows all wavenumbers $k_x,k_y\in[0,\pi/\Delta_x]$.
Denote the Fourier coefficients of $\bm{U}$, $\bm{V}$, and $\bm{b}$ by $\hat{\bm{U}}$, $\hat{\bm{V}}$, and $\hat{\bm{b}}$ respectively.
The Fourier coefficients in the discrete system will evolve according to a linear system of the form
\begin{equation}
\frac{\text{d}}{\text{d}t}\left(\begin{array}{c}\hat{\bm{U}}\\\hat{\bm{V}}\\\hat{\bm{b}}\end{array}\right) = \mat{L}\left(\begin{array}{c}\hat{\bm{U}}\\\hat{\bm{V}}\\\hat{\bm{b}}\end{array}\right)
\end{equation} 
where the entries of the matrix $\mat{L}$ depend on the wavenumbers $k_x$ and $k_y$ and on the particular discretization, either B grid, C grid, or EEN.
Complete details of $\mat{L}$ for each of the three discretizations along with a Matlab code to generate the matrix $\mat{L}$ are given in \cite{BBG18}.

\subsection{Discrete Optimals}
\label{sec:Results}
As noted above, the linear perturbation equations allow specification of an initial condition with a divergent barotropic component, although the barotropic pressure will prevent that component from evolving.
To fully decouple the divergent barotropic component from the dynamics of the system we introduce a matrix $\mat{Q}$ that projects the divergent barotropic component out of the solution.
It is constructed via the auxiliary vector $\hat{\bm{q}}$ which is defined by
\begin{equation}
\hat{\bm{q}} = \left(\begin{array}{c}\bm{1}\hat{\pd{x}}\\\bm{1}\hat{\pd{y}}\\\bm{0}\end{array}\right).
\end{equation}
The notation $\hat{\pd{x}}$ refers to the impact of spatial discretization on the horizontal derivatives; a perfect discretization would have $\hat{\pd{x}} = \imag k_x$.
The matrix 
\begin{equation}
\mat{Q} = \mat{I} - \frac{\hat{\bm{q}}\hat{\bm{q}}^*}{\|\hat{\bm{q}}\|^2}
\end{equation}
is an orthogonal projection matrix that removes the divergent barotropic component of the state.
We completely remove the divergent barotropic component from the dynamics by projecting it out as follows
\begin{equation}
\frac{\text{d}}{\text{d}t}\left(\begin{array}{c}\hat{\bm{U}}\\\hat{\bm{V}}\\\hat{\bm{b}}\end{array}\right) = \mat{LQ}\left(\begin{array}{c}\hat{\bm{U}}\\\hat{\bm{V}}\\\hat{\bm{b}}\end{array}\right).\label{eqn:LQ}
\end{equation}

The discrete energy evolution is easily obtained from (\ref{eqn:LQ})
\begin{equation}
\frac{\text{d}}{\text{d}t} E_{\text{discrete}} = \left(\begin{array}{c}\hat{\bm{U}}\\\hat{\bm{V}}\\\hat{\bm{b}}\end{array}\right)^*\left(\frac{\mat{QL}^*+\mat{LQ}}{2}\right)\left(\begin{array}{c}\hat{\bm{U}}\\\hat{\bm{V}}\\\hat{\bm{b}}\end{array}\right)
\end{equation}
where
\begin{equation}
E_{\text{discrete}}=\frac{1}{2}\left(\begin{array}{c}\hat{\bm{U}}\\\hat{\bm{V}}\\\hat{\bm{b}}\end{array}\right)^*\left(\begin{array}{c}\hat{\bm{U}}\\\hat{\bm{V}}\\\hat{\bm{b}}\end{array}\right).
\end{equation}
The superscript $^*$ denotes the complex conjugate transpose, and the matrix $\mat{Q}$ is Hermitian.
Optimal perturbations are obtained by taking the derivative of the energy with respect to the state variables and setting it proportional to the derivative of the energy tendency with respect to the state variables.
The result is that optimal perturbations are with unit energy are unit eigenvectors of $\mat{QL}^*+\mat{LQ}$, and the associated instantaneous energy growth rates are the eigenvalues.

We next compute the growth rates of the optimal perturbations for the three spatial discretization methods and compare the results to the analytical results from section \ref{sec:Exact}.
In all cases the results were converged with respect to vertical resolution using $N_z=100$.
Also, as in the continuous problem, we observed no dependence on the Richardson number in any of the experiments, so results are presented for $\eps=1$.

We compute results at three resolutions: $\Delta_x = 1/2$, $\Delta_x=1/5$, and $\Delta_x=1/10$.
The nondimensional unit of length is $NH/f$, but the quasigeostrophic deformation radius for this configuration is $NH/(\pi f)$ \cite{RYG16};
the three resolutions correspond to $2/\pi\approx 0.64$, $5/\pi\approx1.59$, and $10/\pi\approx3.18$ grid points per deformation radius.
The grid scale at the lowest resolution is too coarse to represent eddies on the scale of the deformation radius, the middle resolution is `eddy-permitting,' and the highest resolution is `eddy-resolving' in the sense of being able to accurately represent the largest eddies.\\

\begin{figure*}
  \includegraphics[width=\textwidth]{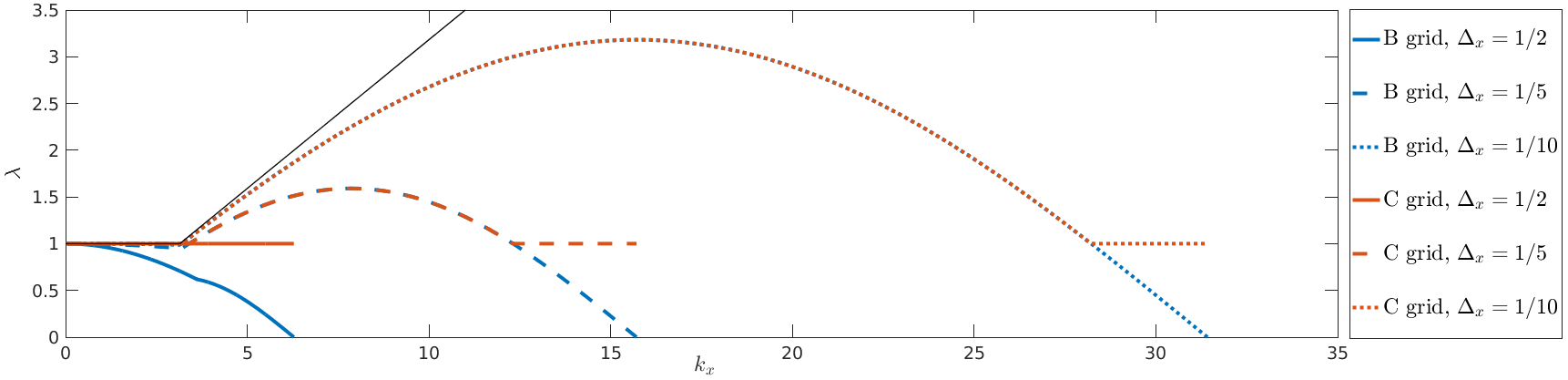}
\caption{Growth rates of instantaneous optimals with no $y$ dependence as a function of wavenumber $k_x$ for the two discretizations -- B grid (blue) and C grid (red) -- at three resolutions: $\Delta_x=1/2$ (solid), $\Delta_x=1/5$ (dashed), $\Delta_x=1/10$ (dotted). (Recall that the two C grid discretizations are identical for perturbations with no $y$ dependence.) The analytical solution is shown as a thin black line.}
\label{fig:Fig2}       
\end{figure*}

We first present results of the three discretizations for perturbations with no $y$ dependence.
As noted in \cite{BBG18}, the linear perturbation equations for the vector-invariant EEN discretization are the same as for the flux-form C grid discretization for perturbations with no $y$ dependence, so the only difference is between B grid and C grid.
The optimal growth rates for the two grids at three resolutions are shown in figure \ref{fig:Fig2}.
In every case the growth rate of the optimal perturbation in the discrete system is too low, and the results are all accurate at large scales (small $k_x$).

At the lowest resolution, the B grid produces poor results.
The C grid methods have quite accurate growth rates $\lambda\approx 1$ but fail to resolve the transition from geostrophic baroclinic perturbations to ageostrophic shear-driven perturbations at $k_x=\pi$.

At the two higher resolutions, the B and C grid methods are remarkably similar except at the smallest scales, where the C grid methods revert to the geostrophic baroclinic growth rate of $\lambda=1$, while the B grid methods have growth rates going to 0.
At intermediate scales, the methods both begin to track the ageostrophic shear-driven perturbations that have $\lambda=k_x/\pi$, but the growth rates become significantly worse (too small) as $k_x$ increases.\\

\begin{figure*}
  \includegraphics[width=\textwidth]{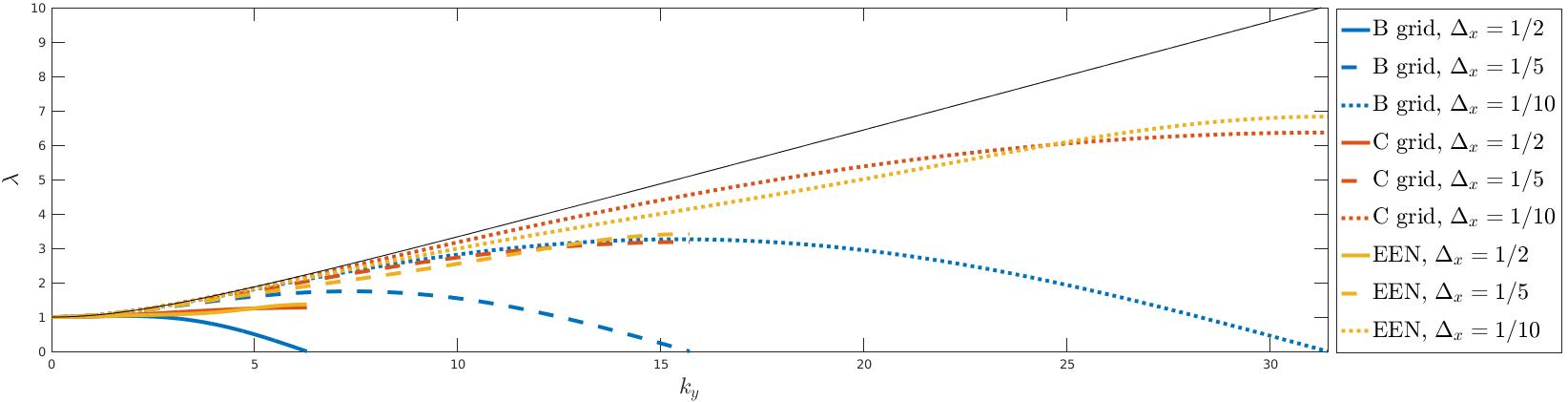}
\caption{Growth rates of instantaneous optimals with no $x$ dependence as a function of wavenumber $k_y$ for the three discretizations: B grid (blue), energy-conserving C grid (red), and energy and enstrophy conserving (EEN) C grid (yellow). Line style indicates grid size: $\Delta_x=1/2$ (solid), $\Delta_x=1/5$ (dashed), $\Delta_x=1/10$ (dotted). The thin black line is the analytical solution.}
\label{fig:Fig3}       
\end{figure*}

We next present results of the three discretizations for perturbations with no $x$ dependence.
The optimal growth rates for the two grids at three resolutions are shown in figure \ref{fig:Fig3}.
At every resolution, the B grid method tracks the correct optimal growth rate for a range of large scales (small $k_y$); the growth rate becomes too weak at larger $k_y$ and finally drops to 0 at the Nyquist wavenumber $\pi/\Delta_x$.

The C grid methods are quite similar to each other, and they are significantly more accurate for perturbations with no $x$ dependence than for perturbations with no $y$ dependence. 
They are also both significantly more accurate than the B grid method at all resolutions.
A key difference compared to the B grid is that the growth rate at the Nyquist scale is nonzero in the C grid methods, which allows the growth rates to remain accurate over a wider range of scales.
For example, at the middle, eddy-permitting resolution the C grid methods are approximately as accurate as the highest resolution B grid method.
The EEN method has slightly better (higher) growth rates than the energy conserving C grid method at the smallest scales and slightly worse (lower) growth rates at intermediate scales.

\section{Discussion \& Conclusions}
\label{sec:Conclusions}
The first result of this paper is analytical expressions for instantaneous optimals in the non-geostrophic Eady problem.
The most basic property of these solutions is that neither their spatial structure nor their growth rates depend in any way on the Richardson number, which is significantly different from the behavior of normal-mode (i.e.~exponential) instability for this problem \cite{Stone66,Stone70,Grooms15,Vallis17}.

For perturbations with no $y$ dependence, the exponentially-unstable modes are confined to a regime of wavenumbers $k_x$ between 0 and some high-wavenumber cutoff.
In contrast, the instantaneous optimals with no $y$ dependence are unstable at all scales.
Optimal perturbations whose growth is based on baroclinic conversion have a fixed growth rate at all scales, while perturbations that grow based on shear production have growth rates linearly increasing with wavenumber.
Kalashnik and Chkhetiani \cite{KC18} computed optimal perturbations with no $y$ dependence and with special restrictions on the potential vorticity in the perturbations for the quasigeostrophic Eady problem.
Their results are broadly similar to our results for perturbations with purely baroclinic growth, except that their growth rates decay as the wavenumber increases.
Farrell and Ioannou \cite{FI96} computed instantaneous optimal perturbations in the discretized quasigeostrophic Eady problem, allowing arbitrary potential vorticity and no $y$ dependence.
Their growth rates appeared flat at high wavenumbers, similar to the results here, but went to zero at small wavenumbers.

For perturbations with no $x$ dependence, exponentially-unstable modes only exist for small-enough Richardson numbers and exist for all wavenumbers above some low-wavenumber cutoff.
In contrast, the instantaneous optimals with no $x$ dependence are unstable at all scales; the growth rate is nonzero at large scales and increases linearly in the small-scale limit.
At large scales, the growth is dominated by baroclinic production, while at small scales the growth is dominated by shear production.
Growing optimal perturbations with both $x$ and $y$ dependence exist for every wavenumber, and the growth rates increase linearly in the limit of small scales.\\

Our goal in studying instantaneous optimals is to learn something about fully nonlinear baroclinic turbulence.
A naive interpretation of our results might suggest that eddy energy in baroclinic turbulence is sustained primarily by shear production at small scales, since growth rates are highest at small scales and the associated optimal perturbations are driven by shear production of perturbation energy.
However, the existence of an instantaneous optimal does not guarantee that it will be strongly excited by turbulent fluctuations, as found by DelSole in a different setting \cite{DelSole04}.
The small-scale optimals associated with shear production may or may not have a strong impact on baroclinic turbulence; it remains to be seen whether they are strongly excited in fully nonlinear baroclinic turbulence.

What seems more relevant to baroclinic turbulence is the behavior of the purely-baroclinic geostrophic optimals that have no $y$ dependence, viz.~that the growth of optimal perturbations is independent of horizontal scale.
Baroclinic turbulence driven by Eady-type background shear excites small-scale features more strongly than baroclinic turbulence driven by other shear profiles \cite{TMHS11,RMCM12,CRKM16}, but the explanation for this cannot be found in linear exponential instability, where small-scale modes are all stable.
Part of the explanation may instead be found in the fact that the geostrophic optimals with no $y$ dependence have growth rates that are independent of wavenumber, implying that small-scale modes can, in principle, extract energy from the mean flow via baroclinic conversion as efficiently as large-scale modes.\\

The second result of this paper is the behavior of instantaneous optimals in spatially-discretized systems using discretization methods that are common to many ocean models \cite{Bryan69,Griffies04,HB99,Madec08,POP}.
The three spatial discretizations considered are the energy conserving Arakawa B grid scheme used by the Parallel Ocean Program \cite{POP}, the energy conserving Arakawa C grid scheme used by the sixth version of the Modular Ocean Model \cite{Griffies04} and in the MITgcm \cite{MAHPH97}, and an energy and enstrophy conserving Arakawa C grid scheme (EEN) available in the Nucleus for European Modelling of the Ocean \cite{Madec08}.
The B grid scheme is the worst among the three, producing growth rates that are too weak and are often significantly worse that the C grid schemes.
The two C grid schemes are extremely similar, with the EEN scheme having slightly better results at the smallest scales and the energy-conserving scheme having slightly better results at intermediate scales.
Both C grid schemes produce growth rates that are too weak at small scales, but, unlike the B grid scheme, the growth rates never go to zero.

These results are markedly different from the results on exponentially-growing unstable modes for the spatially-discrete linear system obtained in \cite{BBG18}.
The authors found in \cite{BBG18} that the C grid methods have rapidly-growing spurious instabilities at small scales.
The results here, being more relevant to fully nonlinear baroclinic turbulence as explained in the introduction, suggest that the spurious exponential instabilities on the C grid are of limited importance for the nonlinear dynamics.
The conclusion here is opposed to that of \cite{BBG18}: we expect the C grid methods to both be more accurate than the B grid for nonlinear baroclinic turbulence, at least insofar as energetic interaction with the mean flow is concerned.
Naturally, nonlinear interactions among turbulent fluctuations are also impacted by the spatial discretization scheme, but our investigation sheds no light on these nonlinear interactions.\\

The ocean mesoscale eddy field is widely held to be energized primarily by geostrophic extraction of large-scale potential energy \cite{FW09}.
Since the ocean mesoscale is strongly nonlinear, the relevance of linear baroclinic stability analysis is limited.
As argued in the introduction, following \cite{DelSole04}, instantaneous optimals are relevant both to linear stability analysis and to analysis of eddy-mean energy exchange in fully-nonlinear turbulence.
The instantaneous optimals that are connected to the baroclinic source of ocean mesoscale eddy energy are presumably the geostrophic, baroclinic optimals at scales near the deformation radius and larger.
These optimals are not as well represented on the B grid as they are on the C grid; in particular, the growth rates are too weak. 
We conclude that the B grid is less-well adapted to modeling the baroclinic production of ocean mesoscale eddy energy when the grid scale is near the eddy scale (both B and C grids are accurate for high-resolution grids).
More specifically, our analysis suggests that the ocean mesoscale eddy field generated by a model with a B grid discretization at intermediate resolution will be less energetic than it should be, not only because of the effects of turbulent viscosity parameterizations but also because of an incorrect representation of baroclinic eddy energy production.


\bibliographystyle{spmpsci}      

\begin{thebibliography}{10}
\providecommand{\url}[1]{{#1}}
\providecommand{\urlprefix}{URL }
\expandafter\ifx\csname urlstyle\endcsname\relax
  \providecommand{\doi}[1]{DOI~\discretionary{}{}{}#1}\else
  \providecommand{\doi}{DOI~\discretionary{}{}{}\begingroup
  \urlstyle{rm}\Url}\fi

\bibitem{AL77}
Arakawa, A., Lamb, V.R.: Computational design of the basic dynamical processes
  of the {UCLA} general circulation model.
\newblock Methods in computational physics \textbf{17}, 173--265 (1977)

\bibitem{AL81}
Arakawa, A., Lamb, V.R.: A potential enstrophy and energy conserving scheme for
  the shallow water equations.
\newblock Mon.~Weather Rev. \textbf{109}(1), 18--36 (1981)

\bibitem{AM88}
Arakawa, A., Moorthi, S.: Baroclinic instability in vertically discrete
  systems.
\newblock J.~Atmos.~Sci. \textbf{45}(11), 1688--1708 (1988)

\bibitem{BBG18}
Barham, W., Bachman, S., Grooms, I.: Some effects of horizontal discretization
  on linear baroclinic and symmetric instabilities.
\newblock Ocean Model. \textbf{125}, 106--116 (2018)

\bibitem{BW88}
Bell, M.J., White, A.A.: Spurious stability and instability in n-level
  quasi-geostrophic models.
\newblock J.~Atmos.~Sci. \textbf{45}(11), 1731--1738 (1988)

\bibitem{BW17}
Bell, M.J., White, A.A.: Analytical approximations to spurious short-wave
  baroclinic instabilities in ocean models.
\newblock Ocean Model. \textbf{118}, 31--40 (2017)

\bibitem{BB88}
B{\"o}berg, L., Br{\"o}sa, U.: Onset of turbulence in a pipe.
\newblock Z.~Naturforsch.~ \textbf{43}(8-9), 697--726 (1988)

\bibitem{Bryan69}
Bryan, K.: A numerical method for the study of the circulation of the world
  ocean.
\newblock J.~Comput.~Phys. \textbf{4}(3), 347--376 (1969)

\bibitem{CRKM16}
Capet, X., Roullet, G., Klein, P., Maze, G.: Intensification of upper-ocean
  submesoscale turbulence through charney baroclinic instability.
\newblock J.~Phys.~Ocean. \textbf{46}(11), 3365--3384 (2016)

\bibitem{DelSole04}
DelSole, T.: The necessity of instantaneous optimals in stationary turbulence.
\newblock J.~Atmos.~Sci. \textbf{61}(9), 1086--1091 (2004)

\bibitem{DlSMB17}
Ducousso, N., Le~Sommer, J., Molines, J.M., Bell, M.: {Impact of the
  “Symmetric Instability of the Computational Kind” at mesoscale-and
  submesoscale-permitting resolutions}.
\newblock Ocean Model. \textbf{120}, 18--26 (2017)

\bibitem{Eady49}
Eady, E.T.: Long waves and cyclone waves.
\newblock Tellus \textbf{1}(3), 33--52 (1949)

\bibitem{Farrell84}
Farrell, B.: Modal and non-modal baroclinic waves.
\newblock J.~Atmos.~Sci. \textbf{41}(4), 668--673 (1984)

\bibitem{Farrell85}
Farrell, B.: Transient growth of damped baroclinic waves.
\newblock J.~Atmos.~Sci. \textbf{42}(24), 2718--2727 (1985)

\bibitem{Farrell89}
Farrell, B.F.: Optimal excitation of baroclinic waves.
\newblock J.~Atmos.~Sci. \textbf{46}(9), 1193--1206 (1989)

\bibitem{FI96}
Farrell, B.F., Ioannou, P.J.: Generalized stability theory. {P}art {I}:
  Autonomous operators.
\newblock J.~Atmos.~Sci. \textbf{53}(14), 2025--2040 (1996)

\bibitem{FW09}
Ferrari, R., Wunsch, C.: Ocean circulation kinetic energy: Reservoirs, sources,
  and sinks.
\newblock Annu.~Rev.~Fluid Mech. \textbf{41} (2009)

\bibitem{Griffies04}
Griffies, S.M.: Fundamentals of ocean climate models.
\newblock Princeton University Press (2004)

\bibitem{Grooms15}
Grooms, I.: Submesoscale baroclinic instability in the {Balance Equations}.
\newblock J.~Fluid Mech. \textbf{762}, 256--272 (2015)

\bibitem{HB99}
Haidvogel, D.B., Beckmann, A.: Numerical ocean circulation modeling, vol.~2.
\newblock World Scientific (1999)

\bibitem{HKRB83}
Hollingsworth, A., K{\aa}llberg, P., Renner, V., Burridge, D.M.: An internal
  symmetric computational instability.
\newblock Q.~J.~Roy.~Meteor.~Soc. \textbf{109}(460), 417--428 (1983)

\bibitem{KC18}
Kalashnik, M.V., Chkhetiani, O.: An analytical approach to the determination of
  optimal perturbations in the {Eady} model.
\newblock J.~Atmos.~Sci. \textbf{75}, 2741--2761 (2018)

\bibitem{lSPTMB09}
Le~Sommer, J., Penduff, T., Theetten, S., Madec, G., Barnier, B.: How momentum
  advection schemes influence current-topography interactions at eddy
  permitting resolution.
\newblock Ocean Model. \textbf{29}(1), 1--14 (2009)

\bibitem{Madec08}
Madec, G.: {NEMO} ocean engine.
\newblock Note du P\^ole de mod{\'e}lisation, Institut Pierre-Simon Laplace
  (IPSL), France, No 27, ISSN No 1288-1619 (2008)

\bibitem{MAHPH97}
Marshall, J., Adcroft, A., Hill, C., Perelman, L., Heisey, C.: A finite-volume,
  incompressible {N}avier-{S}tokes model for studies of the ocean on parallel
  computers.
\newblock J.~Geophys.~Res.-Oceans \textbf{102}(C3), 5753--5766 (1997)

\bibitem{RYG16}
Rocha, C.B., Young, W.R., Grooms, I.: On {Galerkin} approximations of the
  surface active quasigeostrophic equations.
\newblock J.~Phys.~Ocean. \textbf{46}(1), 125--139 (2016)

\bibitem{RMCM12}
Roullet, G., McWilliams, J.C., Capet, X., Molemaker, M.J.: Properties of steady
  geostrophic turbulence with isopycnal outcropping.
\newblock J.~Phys.~Ocean. \textbf{42}(1), 18--38 (2012)

\bibitem{Schmid07}
Schmid, P.J.: Nonmodal stability theory.
\newblock Annu.~Rev.~Fluid Mech. \textbf{39}, 129--162 (2007)

\bibitem{SH01}
Schmid, P.J., Henningson, D.S.: Stability and transition in shear flows.
\newblock Springer (2001)

\bibitem{Smith07}
Smith, K.S.: {The geography of linear baroclinic instability in Earth's
  oceans}.
\newblock J.~Marine Res. \textbf{65}(5), 655--683 (2007)

\bibitem{POP}
Smith, R., Jones, P., Briegleb, B., Bryan, F., Danabasoglu, G., Dennis, J.,
  Dukowicz, J., Eden, C., Fox-Kemper, B., Gent, P., et~al.: The parallel ocean
  program ({POP}) reference manual.
\newblock Los Alamos National Lab Technical Report \textbf{141} (2010)

\bibitem{Stone66}
Stone, P.H.: On non-geostrophic baroclinic stability.
\newblock J.~Atmos.~Sci. \textbf{23}(4), 390--400 (1966)

\bibitem{Stone70}
Stone, P.H.: On non-geostrophic baroclinic stability: Part {II}.
\newblock J.~Atmos.~Sci. \textbf{27}(5), 721--726 (1970)

\bibitem{TE05}
Trefethen, L.N., Embree, M.: Spectra and Pseudospectra of nonnormal matrices
  and operators.
\newblock Princeton University Press (2005)

\bibitem{TMHS11}
Tulloch, R., Marshall, J., Hill, C., Smith, K.S.: Scales, growth rates, and
  spectral fluxes of baroclinic instability in the ocean.
\newblock J.~Phys.~Ocean. \textbf{41}(6), 1057--1076 (2011)

\bibitem{Vallis17}
Vallis, G.K.: Atmospheric and Oceanic Fluid Dynamics: Fundamentals and
  Large-scale Circulation, {Second} edn.
\newblock Cambridge University Press (2017)

\end{thebibliography}

\end{document}